\documentclass[12pt]{article}
\usepackage{amsfonts}
\usepackage{pstricks}
\usepackage{pst-node}
\usepackage{epsfig}
\usepackage{amsmath,amssymb,latexsym}
\usepackage{xcolor}
\usepackage{hyperref}

\usepackage{import}

\usepackage{enumerate}

\usepackage{physics}

\usepackage[lmargin=1.0in, rmargin=1.0in, tmargin=1.0in, bmargin=1.0in]{geometry}

\usepackage{cancel}

\usepackage{empheq}

\usepackage{comment}

\usepackage{setspace}
\begin{document}



\title{\bf Weyl-Lorentz-$U(1)$-invariant symmetric teleparallel gravity in three dimensions}
\author{ Muzaffer Adak$^{1,2}$, Nese Ozdemir$^{2}$, Caglar Pala$^{1,3,4}$ \\
  {\small $^1$Department of Physics, Faculty of Science, Pamukkale University, Denizli, Türkiye} \\
  {\small $^2$Department of Physics,  Istanbul Technical University, İstanbul, Türkiye} \\
  {\small $^3$ Laboratory of Theoretical Physics, Institute of Physics, University of Tartu, Tartu, Estonia} \\
  {\small $^4$ Department of Physics, Faculty of Science,  Erciyes University, Kayseri, Türkiye}\\
  {\small {\it E-mail:} {\blue madak@pau.edu.tr, nozdemir@itu.edu.tr, caglar.pala@gmail.com}}
 }
  \vskip 1cm
\date{\today}
\maketitle

 \thispagestyle{empty}

\begin{abstract}
 \noindent
We consider a Weyl-Lorentz-$U(1)$-invariant gravity model written in terms of a scalar field, electromagnetic field and nonmetricity without torsion and curvature, the so-called symmetric teleparallel geometry, in three dimensions. Firstly, we obtain variational field equations from a Lagrangian. Then, we find some classes of circularly symmetric rotating solutions by making {\it only} a metric ansatz. The coincident gauge of symmetric teleparallel spacetime allows us for doing so. \\


\noindent PACS numbers: 04.50.Kd, 11.15.Kc, 02.40.Yy \\ 
 {\it Keywords}: Modified gravity, non-Riemannian geometry, nonmetricity, calculus of variation, Weyl (scale) invariance

\end{abstract}

\section{Introduction}

Einstein's theory of gravity, General Relativity (GR), is described by field equations containing the second order derivatives of metric components. Those equations are derived by a calculus of variation from the Einstein-Hilbert action integral. GR is in perfect agreement with the observational results in the solar system. On the other hand, recently the observed results of change of linear velocities of stars in the outer arms of disk-shaped galaxies with distance from the center of galaxy do not agree with the predictions of GR. This mismatch is called the dark matter problem \cite{trimble-1987}. In addition, when the velocities of galaxies are measured, it is observed that they increase instead of decrease, which is the prediction of GR. This contradiction is called the cosmic acceleration problem. Today, we explain this observation with hypothetical dark energy, which we don't know its content \cite{peebles-ratra-2003}. Furthermore, the vain efforts to quantise GR \cite{kraichnan-1955},\cite{deser-1970},\cite{isham_1981} give strong signals that it needs to be altered. For these reasons, the search for alternative gravity models is a very important and hot research field from point of view of theoretical physics. The search for a new theory of gravity free from the problems mentioned here can be done in different ways. However, we will try to do this with the help of non-Riemannian geometries in this paper. While doing this, we will work in a spacetime geometry where full curvature and torsion are zero, but the nonmetricity is not zero. Such geometries are known as symmetric teleparallel spacetime or Minkowski-Weyl spacetime. More mathematical motivations for the choice of such geometry will be given in the subsequent section.

On the other hand, 3-dimensional gravity models have attracted a lot of attention in recent years, because GR in (2+1) dimensions has no propagating degrees of freedom and quantum gravity studies give promising results in 3-dimensional spacetimes. For example, Banados-Teitelboim-Zanelli (BTZ) found a 3-dimensional black hole solution to Einstein's equations with negative cosmological constant \cite{banados-btz-1992}. That has revealed interesting properties at both the classical and quantum levels, and also has some common behaviors with the 4-dimensional Kerr black hole \cite{carlip-2005}. In fact, it is the presence of this BTZ black hole that makes 3-dimensional gravity an attractive field of theoretical physics studies. Meanwhile, if the odd-parity Chern-Simons term is added to the Einstein-Hilbert action, a 3-dimensional theory of propagating gravity, the so-called Topologically Massive Gravity (TMG), is obtained \cite{deser_jackiw_temp_1982}, \cite{deser_jackiw_temp_1982_2}. Later, it was demonstrated by holographic methods that the TMG model has a unitary or bulk versus boundary clash problem \cite{kraus-larsen-2006}. To eliminate this problem, a new massive gravity (NMG) theory has been proposed by adding even-parity quadratic curvature terms and discarding the odd-parity topological Chen-Simons term of TMG \cite{bergshoeff-2009-prl}. However, it has been seen that the NMG model cannot fully solve the problems encountered in the TMG model. Accordingly, Generalized Massive Gravity (GMG) model was introduced by adding the odd-parity Chern-Simons term back to the NMG action in the hope that the bulk versus boundary problem would be solved \cite{bergshoeff-2009-prd}. Then, the Minimal Massive Gravity (MMG) model \cite{bergshoeff-2014}, obtained by adding torsion and an axial field to the TMG model, solved the bulk-boundary clash problem in certain ranges of the parameters \cite{arvanitakis}. Although such 3-dimensional models of gravity have no direct effects for real-world gravity, they are likely to have applications in solid state physics \cite{bergshoeff-2018-prl}. For further read on 3-dimensional gravity models one may consult for \cite{hortacsu-nese-2008}-\cite{adak-ozdemir-2023}. Despite the above efforts on 3-dimensional gravitational researches, the formulation of a quantum
theory of gravity is definitely still an open problem. Thus, it is worthwhile to perform further effort on 3-dimensional scenarios.

\section{Mathematical motivations}

After the brief and fast summary of literature on why modified theories of gravity and 3-dimensional spacetime have to be studied, in this section we give more mathematical motivations on our work. GR is formulated in the language of exterior algebra in 4-dimensional spacetimes as follows
 \begin{align}
     \widetilde{G}_a := \frac{1}{2} \widetilde{R}^b{}_c \wedge *(e_a \wedge e_b \wedge e^c) = \kappa \widetilde{\tau}_a[mat],
 \end{align}
where $e^a$ is the $g$-orthonormal coframe (or orthonormal basis 1-form), $\widetilde{R}^a{}_b$ is the Riemann curvature 2-form which is expressed uniquely in terms $g$-orthonormal coframe, $*$ represents Hodge dual map, $\kappa$ is a coupling constant, $\widetilde{\tau}_a[mat]$ denotes energy-momentum 3-form of matter and $\widetilde{G}_a$ is the Einstein tensor 3-form. It is well known that one can obtain $e^a$ and $\widetilde{R}^a{}_b$ from the metric tensor, $g=\eta_{ab} e^a \otimes e^b$ where $\eta_{ab}$ is the Minkowski metric with the signature $(-,+,+,+)$. In four dimensions Einstein tensor 3-form has 16 components. On the other hand, the Riemannian curvature 2-form has 20 independent components (36 from $\widetilde{R}^a{}_{b}$ minus 16 from the Bianchi identity, $\widetilde{R}^a{}_{b} \wedge e^b=0$). Thus, in vacuum, i.e., $\widetilde{\tau}_a[mat]=0$, though all components of the Einstein tensor vanish, some components of $\widetilde{R}^a{}_b$ may still live and then gravitational waves are allowed in an empty spacetime. 

If one does the similar analysis in a three-dimensional spacetime, it is seen that there are 9 components of $\widetilde{G}_a$ 2-form and 6 independent components of $\widetilde{R}^a{}_b$ 2-form. Consequently, as the Einstein tensor vanishes, all components of $\widetilde{R}^a{}_{b}$ must also be zero. It means that there can not be gravitational waves in vacuum. Correspondingly, the bare Einstein's general relativity is not a dynamical theory in 3-dimensions. Therefore, there is a wide literature on 3-dimensional modified theories of gravity \cite{banados-btz-1992}-\cite{adak-ozdemir-2023}.

One way of modifying the Einstein's theory of gravity is to go beyond the Riemannian geometry. In this work we render the symmetric teleparallel geometry which is characterized by only the nonmetricity tensor \cite{nester-1999}-\cite{caglar_ozcan_muzo_2022}. One can read \cite{adak-dereli-2023} in order to catch the literature on the teleparallel geometries consisting of the metric (Weitzenböck) teleparallel geometry described by only torsion, the symmetric teleparallel geometry described by only nonmetricity and the general teleparallel geometry described by torsion and nonmetricity. Now, we count the numbers of components of equations and nonmetricity in 3 dimensions in order to see if there is a similar feature as in the Einstein's equation. The field equation for the bare symmetric teleparallel gravity  (STPG) in vacuum is written in the equation (\ref{eq:field-eqn-combined-stpg}), with $\Lambda=0$, 
 \begin{align}
       \iota_b D\Sigma^b{}_a  +  \tau_a  = 0 .  \label{eq:stpg-field-eqn}
 \end{align}
Here we know that the 2-forms $\tau_a$ and $\Sigma^b{}_a$ contain the metric-orthonormal coframe $e^a$ and the nonmetricity 1-form $Q_{ab}$. As seen explicitly in the forthcoming sections $e^a$ is determined by metric in general and $Q_{ab}$ could be computed in terms of metric in the coincident gauge. In spite of that the bare STPG equation has 9 components, $Q_{ab}$ has 18 components. Therefore, we could expect that STPG might accommodate also theories with non-trivial local dynamics\footnote{Another heuristic argument can be based on symmetry considerations. The metric tensor in 3 dimensions has 6 independent components, but due to the diffeomorphism invariance of Einstein's theory, we are left with $6-2\times 3 =0$ dynamical degrees of freedom. However, a generic STPG theory breaks the symmetry, rendering (some of) degrees of freedom from gauge to dynamical.}. Even in 2 dimensions, in which GR is trivial, STPG is a dynamical theory in vacuum \cite{adak2008}.

We generally consider STPG models which are invariant under the Lorentz transformations which form Lorentz group, $SO(1,2)$. Meanwhile, $SO(1,2)$ is doubly folded by $Spin(1,2)$ group that is generated by the even subalgebra, $Cl_+(1,2)$, of the Clifford algebra. Furthermore, $Cl_+(1,2)$ is spanned by the set $\{1, \sigma_{ab}\}$. As the element, $\sigma_{ab}=-\sigma_{ba}$, generates the Lorentz group by the exponentiation $e^{\sigma_{ab}\theta^{ab}(x)/2} \in SO(1,2)$, the unity generates the Weyl (scale) group, $e^{1 \psi(x)} \in \mathcal{W}$, where $\theta_{ab}(x)$ and $\psi(x)$ are the transformation parameters. Consequently, we think that the complete gauge group for STPG models should be $SO(1,2)\otimes \mathcal{W}$ \cite{caglar_ozcan_muzo_2022}. This is the mathematical motivation for the inclusion of a scalar field to a model of STPG.

Under a $SO(1,2)$ transformation the metric, $g=\eta_{ab} e^a \otimes e^b$, and the affine connection 1-form, $\omega^a{}_b$, transform as follows
  \begin{subequations}
     \begin{align}
     e^{a'} &= L^{a'}{}_a e^a \quad \text{and} \quad \eta_{a'b'} = L^a{}_{a'} L^b{}_{b'} \eta_{ab}, \\
     \omega^{a'}{}_{b'} &= L^{a'}{}_a L^b{}_{b'} \omega^a{}_b + L^{a'}{}_{a}dL^a{}_{b'},
 \end{align}
 \end{subequations}
where $L^a{}_{a'}(x) , L^{a'}{}_a(x) \in SO(1,2)$. On the other hand, under a Weyl transformation they behave as
  \begin{subequations}
     \begin{align}
     \hat{e}^{a} &= e^\psi e^a \quad \text{and} \quad \hat{\eta}_{ab} =  \eta_{ab}, \\
     \hat{\omega}^{a}{}_{b} &= \omega^a{}_b - \delta^a_b d\psi , \label{eq:connect-weyl}
 \end{align}
 \end{subequations} 
where $e^\psi(x) \in \mathcal{W}$. In fact, one can choose $\hat{\omega}^{a}{}_{b} = \omega^a{}_b$ instead of (\ref{eq:connect-weyl}), because both leave the curvature, $R^a{}_b := d\omega^a{}_b + \omega^a{}_c \wedge \omega^c{}_b$, invariant. However, since we want to stay in the symmetric teleparallel geometry before and after a Weyl transformation we prefer the rule (\ref{eq:connect-weyl}) rather than $\hat{\omega}^{a}{}_{b} = \omega^a{}_b$. It is worthy to notice that the other option yields some torsion after a Weyl transformation \cite{adak-ozdemir-2023}. In general, the simplest geometry which is invariant under a Weyl transformation is the symmetric teleparallel one \cite{caglar_ozcan_muzo_2022}. That is one of reasons for adhering the symmetric teleparallel geometry.  

In addition to these motivations, we couple the electromagnetic field to our model, as it is among our future projects to apply the mathematical techniques we have developed here to our photonic crystal studies \cite{yuksel-berberoglu-2022a}-\cite{oguz-karakilinc-2022b}. Consequently, firstly we consider a modified theory of gravity, the so-called STPG, formulated in 3-dimensional symmetric teleparallel spacetime, then couple minimally the Maxwell field to it and finally add a scalar field to the STPG-Maxwell theory in a non-minimal way for scale invariance in order to gain physical insights about nonmetricity and to see new interactions among nonmetricity, electromagnetic field and scalar field. 

The plan of the work is as follows. Since we will use the language of exterior algebra through the paper in the next section we summarize our notations and conventions, then formulate our theory by writing down a Lagrangian 3-form. After obtaining variational field equations we search a circularly symmetric rotating solutions by using the coincident gauge of the symmetric teleparallel geometry. We especially give all details of how the coincident gauge is used in calculations. Then, we obtain some classes of exact solutions together with discussions of singularity structure for each case. As looking for exact solutions we extensively use the computer algebra system REDUCE \cite{hearn-2004} and its exterior algebra package EXCALC \cite{schrufer-2004}. In the section of Discussion, we collect our results, relate our findings with material science literature and remark on some future projects.

\section{Notations, conventions, definitions}

The triple $\{M,g,\nabla\}$ defines a spacetime where $M$ is three-dimensional orientable and differentiable manifold, $g$ is non-degenerate metric and $\nabla$ is full (or affine) connection \cite{thirring1997}, \cite{frankel2012}. We denote the $g$-orthonormal coframe by $e^a$, then write the metric as $g=\eta_{ab} e^a \otimes e^b$ where $\eta_{ab}$ is the Minkowski metric with the signature $(-,+,+)$. The full connection is determined by the full connection 1-form $\omega^a{}_b$ via the definition $\nabla e^a := - \omega^a{}_b \wedge e^b$. In the language of exterior algebra, $e^a$ is called orthonormal 1-form and the Cartan structure equations are given by nonmetricity 1-form, torsion 2-form and curvature 2-form tensors, respectively,
 \begin{subequations}\label{eq:cartan-ort}
 \begin{align}
     Q_{ab} &:= -\frac{1}{2} D\eta_{ab} = \frac{1}{2} (\omega_{ab} + \omega_{ba}), \label{eq:nonmetric}\\
     T^a &:= De^a = de^a + \omega^a{}_b \wedge e^b, \label{eq:tors}\\
     R^a{}_b &:= D\omega^a{}_b := d \omega^a{}_b + \omega^a{}_c \wedge \omega^c{}_b, \label{eq:curv}
 \end{align}
 \end{subequations}
where $d$ is the exterior derivative, $D$ is the Lorentz-covariant exterior derivative and $\wedge$ is the exterior product. They satisfy the Bianchi identities
  \begin{align}
      DQ_{ab}= \frac{1}{2}(R_{ab}+R_{ba}) , \qquad DT^a = R^a{}_b \wedge e^b , \qquad DR^a{}_b=0 .
  \end{align}

The full connection, $\omega^a{}_b$, can be decomposed uniquely to a Riemannian piece, $\widetilde{\omega}_{ab}(g)$ determined by metric, and a non-Riemannian piece, $\mathrm{L}_{ab}(T,Q)$ determined by torsion and nonmetricity \cite{tucker1995},\cite{hehl1995},\cite{adak2022ijgmmp},
 \begin{align}
     \omega_{ab}=\widetilde{\omega}_{ab} + \mathrm{L}_{ab}, \label{eq:connec-decom}
 \end{align}
where $\widetilde{\omega}_{ab}$ is the anti-symmetric 1-form, the so-called Levi-Civita connection 1-form, 
 \begin{equation}\label{eq:Levi-Civita}
   \widetilde{\omega}_{ab} = \frac{1}{2} \left[ -\iota_a de_b + \iota_b de_a + (\iota_a \iota_b de_c) e^c \right] \qquad \text{or} \qquad \widetilde{\omega}^a{}_b \wedge e^b = -de^a
 \end{equation}
and $\mathrm{L}_{ab}$ is an asymmetric 1-form, the so-called distortion tensor 1-form, 
 \begin{equation}
     \mathrm{L}_{ab}=Q_{ab} + ( \imath_b Q_{ac} - \imath_a Q_{bc} ) e^c + \frac{1}{2} \left[ \iota_a T_b - \iota_b T_a - (\iota_a \iota_b T_c) e^c \right].
 \end{equation}
Here $\iota_a \equiv \iota_{X_a}$ denotes the interior product with respect to the orthonormal base vector $X_a$. In the literature it is common to define the anti-symmetric contortion tensor 1-form, $K_{ab}$, in terms of torsion tensor 2-form
 \begin{align}
     K_{ab} = \frac{1}{2} \left[ \iota_a T_b - \iota_b T_a - (\iota_a \iota_b T_c) e^c \right] \qquad \text{or} \qquad K^a{}_b \wedge e^b = T^a .
 \end{align}
It is worthy to notice that the symmetric part of the affine connection is determined by only nonmetricity, $\omega_{(ab)}= Q_{ab}$, the remainder of $\omega_{ab}$ is the anti-symmetric\footnote{Here $\omega_{(ab)}:= \frac{1}{2} (\omega_{ab}+\omega_{ba}) = Q_{ab}$ is the reason why we adopt the factor $-\frac{1}{2}$ in the definition of nonmetricity (\ref{eq:nonmetric}).}. All the Riemannian quantities will be labelled by a tilde over them in this paper.

Under a Lorentz transformation, $SO(1,2)$, the quantities transform as follows
 \begin{align}
     e^{a'}=L^{a'}{}_a e^a , \quad \eta_{a'b'} = L^a{}_{a'}L^b{}_{b'} \eta_{ab} , \quad \omega^{a'}{}_{b'} = L^{a'}{}_{a} \omega^{a}{}_{b} L^{b}{}_{b'} + L^{a'}{}_{a} d L^{a}{}_{b'}, \quad \phi' = \phi , \quad A' = A \label{eq:lorentz}
 \end{align}
where $\phi$ is a scalar field and $A$ is the Maxwell potential 1-form. Since $\eta_{a'b'} = \eta_{ab} = \text{diag}(-1,+1,+1)$, transformation elements depending on local coordinates, $L^{a'}{}_a(x)$ and $L^{a}{}_{a'}(x)$, form the Lorentz group, and also satisfy $L^{a'}{}_a L^{a}{}_{b'} = L^{a}{}_{a'} L^{a'}{}_{b}  = \delta^a_b$. Similarly, a coordinate dependent Weyl (scale) transformation yields
 \begin{align}
     \hat{e}^{a} = e^{\psi(x)} e^a, \quad \hat{\eta}_{ab} = \eta_{ab} , \quad \hat{\omega}^{a}{}_{b} = \omega^a{}_b - \delta^a_b d\psi(x), \quad \hat{\phi} = e^{-\psi(x)} \phi, \quad \hat{A} = A \label{eq:weyl}
 \end{align}
where the transformation function, $\psi(x)$, is real and $e^{\psi(x)}$ forms the Weyl group, $\mathcal{W}$. The effect of these transformation rules on the Hodge map is as follows
  \begin{align}
      \hat{*} \, \Phi = e^{(3-2p)\psi} *\Phi 
  \end{align}
where $\Phi$ is any $p$-form. Finally under a local $U(1)$ transformation they behave as
 \begin{align}
     e^a \to  e^a, \quad \eta_{ab} \to \eta_{ab} , \quad \omega^{a}{}_{b} \to \omega^a{}_b, \quad \phi \to \phi, \quad A \to  A + df(x) \label{eq:unitary}
 \end{align}
where the transformation function is real and $e^{if(x)} \in U(1)$ together with $i=\sqrt{-1}$. Accordingly, we define the Lorentz-covariant exterior derivative of any $(p,q)$-type tensor-valued exterior form $\mathfrak{T}^{a_1 a_2 \cdots a_p}_{\; \; \; \; b_1 b_2 \cdots b_q }$ below
 \begin{eqnarray}\label{eq:gl-covariant-derivative}
   D \mathfrak{T}^{a_1 a_2 \cdots a_p}_{\; \; \; \; b_1 b_2 \cdots b_q } = d \mathfrak{T}^{a_1 a_2 \cdots a_p}_{\; \; \; \; b_1 b_2 \cdots b_q }
    + \omega^{a_1}{}_c \wedge \mathfrak{T}^{c a_2 \cdots a_p}_{\; \; \; \; b_1 b_2 \cdots b_q } + \cdots + \omega^{a_p}{}_c \wedge \mathfrak{T}^{a_1 a_2 \cdots c}_{\; \; \; \; b_1 b_2 \cdots b_q } \nonumber \\
     - \omega^c{}_{b_1} \wedge \mathfrak{T}^{a_1 a_2 \cdots a_p}_{\; \; \; \; c b_2 \cdots b_q } - \cdots
    - \omega^c{}_{b_q} \wedge \mathfrak{T}^{a_1 a_2 \cdots a_p}_{\; \; \; \; b_1 b_2 \cdots c} .
 \end{eqnarray}
In the calculations we use some abbreviations and identities
 \begin{subequations}
 \begin{align}
     e^{ab\cdots} &:= e^a \wedge e^b \wedge \cdots , \quad
     \iota_{ab \cdots } := \iota_a \iota_b \cdots , \quad
     \partial_a := \iota_a d , \\
     D*e_a &= -Q \wedge *e_a + *e_{ab} \wedge T^b, \\
     D*e_{ab} &= -Q \wedge *e_{ab} + *e_{abc} \wedge T^c, \\
     D*e_{abc} &= -Q \wedge *e_{abc}  ,
 \end{align}
 \end{subequations}
where $*$ is the Hodge map fixed by $*1= \frac{1}{3!} \epsilon_{abc} e^a \wedge e^b \wedge e^c = e^{012}$ in terms of totally antisymmetric epsilon tensor, $\epsilon_{abc}$, and $Q=\eta_{ab}Q^{ab}$ is the nonmetricity trace 1-form.

\section{Scale invariant symmetric teleparallel theory of gravity coupled to electromagnetic field}

Instead of attacking impatiently the most general problem we prefer to approach it in three steps from relatively easier to harder. Therefore, firstly we forget scalar field and electromagnetic field and consider the bare symmetric teleparallel gravity (STPG).

\subsection{STPG in three dimensions}

We formulate STPG by writing the Lagrangian 3-form
  \begin{align} \label{eq:lag-stpg}
     L_1[Q] =& L[Q^2] + \Lambda *1 + T^a \wedge \lambda_a + R^a{}_b \wedge \rho^b{}_a 
 \end{align}
where  $\Lambda$ is cosmological constant, $\lambda_a$ and $\rho^b{}_a$ are Lagrange multiplier 1-forms constraining torsion and curvature to zero, respectively, and $L[Q^2]$ is the even parity quadratic Lagrangian 3-form,
 \begin{align} \label{eq:lag-QQ}
     L[Q^2] =& c_1  Q_{ab} \wedge *Q^{ab} + c_2 \left(Q_{ab} \wedge e^b\right) \wedge *\left(Q^{ac} \wedge e_c\right) + c_3 \left(Q_{ab} \wedge e_c \right) \wedge *\left(Q^{ac} \wedge e^b\right)  \nonumber \\
     & \quad + c_4 Q \wedge *Q   + c_5 (Q \wedge e^b) \wedge * (Q_{ab} \wedge e^a) .
 \end{align}
Here $c_i$, $i=1,2,..,5$, are coupling constants. In order for our theory formulated in terms of exterior algebra to be compared easily with the literature using tensor formulation we firstly write  $Q_{ab}=Q_{abc}e^c$ where $(0,3)$-type $Q_{abc}$ nonmetricity tensor 0-form is symmetric in the first two indices, $Q_{abc}=Q_{(ab)c}$. Correspondingly, $L[Q^2]$ turns out to be,
 \begin{align} \label{eq:lag-QQ2}
     L[Q^2] =& \big\{ (c_1+c_2+c_3) Q_{abc} Q^{abc} -c_2 Q_{abc}Q^{acb} -c_3 Q_{ab}{}^b Q^{ac}{}_c \nonumber \\
     & \qquad \qquad \qquad \qquad  + (c_4+c_5) Q^a{}_{ac} Q_b{}^{bc} -c_5 Q^a{}_{ab} Q^{bc}{}_c \big\} *1 .
 \end{align}
The Lagrangian (\ref{eq:lag-QQ}) or (\ref{eq:lag-QQ2}) is Lorentz invariant, and also equivalent to GR for the values of $c_i$s,
  \begin{align} \label{eq:gr-equiv-values-ci}
      c_1=  \frac{1}{2\kappa} , \quad c_2 =- \frac{1}{\kappa} , \quad c_3 =0 , \quad c_4 = - \frac{1}{2\kappa}, \quad c_5= \frac{1}{\kappa} ,
  \end{align}
where $\kappa$ is coupling constant. These values are consistent with those written just after the equation (54) of \cite{koivisto-Iosifidis2018} and with (31) of  \cite{adak2013ijmpa}. If one wants to see the standard results, it is enough to substitute these values of $c_i$s to the solutions. 

We now will perform calculation of variation on the Lagrangian (\ref{eq:lag-stpg}). We always use the general result from \cite{adak2006},
 \begin{align}
      \delta(\alpha \wedge *\beta) = \delta \alpha \wedge *\beta + \delta \beta \wedge *\alpha - \delta e^a \wedge \big[ (\iota_a\beta) \wedge *\alpha - (-1)^p \alpha \wedge (\iota_a *\beta) \big] 
 \end{align}
where $\alpha$ and $\beta$ are any $p$-forms in $n$ dimensions, ($0 \leq p \leq n$). So, the variations $\delta \lambda_a$, $\delta \rho^b{}_a$, $\delta e^a$ and $\delta \omega^a{}_b$ yield the following field equations, respectively,
 \begin{subequations} \label{eq:equations-stpg}
 \begin{align}
     T^a &= 0 , \label{eq:zero-tors-stpg}\\
     R^a{}_b &= 0, \label{eq:zero-curv-stpg} \\
      \tau_a[Q] + \Lambda *e_a + D\lambda_a &= 0 , \label{eq:field-eqn-coframe-stpg} \\
      \Sigma^b{}_a[Q] +  e^b \wedge \lambda_a + D\rho^b{}_a &= 0, \label{eq:field-eqn-connec-stpg}
 \end{align}
  \end{subequations}
where we dismissed the exact forms. Energy-momentum 2-form of nonmetricity is 
  \begin{align} 
 \tau_a[Q]=  \sum_{i=1}^5 c_i \overset{(i)}{\tau_a}[Q] \label{eq:tauQ-stpg}
 \end{align}
where
  \begin{subequations} \label{eq:enr-mom-nonmet-sqred}
 \begin{align}
    \overset{(1)}{\tau_a}[Q] =& - \left( \iota_a  Q^{bc}  \right) \wedge * Q_{bc} - Q_{bc} \wedge \left( \iota_a* Q^{bc}  \right)  , \\
    \overset{(2)}{\tau_a}[Q] =& - 2Q_{ab} \wedge *\left( Q^{bc} \wedge e_c \right) - \left[ \iota_a\left( Q^{dc} \wedge e_c \right) \right] \wedge *\left( Q_{db} \wedge e^b \right) \nonumber \\
    &   \qquad \qquad + \left( Q_{db} \wedge e^b \right) \wedge \left[ \iota_a*\left( Q^{dc} \wedge e_c \right)\right]  , \\
    \overset{(3)}{\tau_a}[Q] =& - 2Q^{bc} \wedge *\left( Q_{ac} \wedge e_b \right) - \left[ \iota_a\left( Q^{dc} \wedge e^b \right) \right] \wedge *\left( Q_{db} \wedge e_c \right) \nonumber \\
    & \qquad \qquad + \left( Q_{db} \wedge e_c \right) \wedge \left[ \iota_a*\left( Q^{dc} \wedge e^b \right)\right]  , \\
   \overset{(4)}{\tau_a}[Q] =& - \left( \iota_a  Q \right) \wedge * Q - Q \wedge \left( \iota_a* Q  \right)  , \\
    \overset{(5)}{\tau_a}[Q] =& - Q \wedge *\left( Q_{ab} \wedge e^b \right) - Q_{ab} \wedge *\left( Q \wedge e^b \right) - \left[ \iota_a\left( Q_{bc} \wedge e^c \right) \right] \wedge *\left( Q \wedge e^b \right) \nonumber \\
    & \qquad \qquad + \left( Q \wedge e^b \right) \wedge \left[ \iota_a*\left( Q_{bc} \wedge e^c \right)\right]  .
 \end{align}
 \end{subequations}
Angular (hyper) momentum 2-form of nonmetricity is 
  \begin{align}  
 \Sigma^b{}_a[Q]= \sum_{i=1}^5 c_i \overset{(i)}{\Sigma^b{}_a}[Q] \label{eq:SigmaQQ-stpg}
 \end{align}
where
  \begin{subequations} \label{eq:ang-mom-nonmet-sqred}
  \begin{align}
       \overset{(1)}{\Sigma^b{}_a}[Q] =& 2 *Q^b{}_a  ,\\
      \overset{(2)}{\Sigma^b{}_a}[Q] =&  e^b \wedge *\left( Q_{ac} \wedge e^c \right) +  e_a \wedge *\left( Q^{bc} \wedge e_c \right)  ,\\
      \overset{(3)}{\Sigma^b{}_a}[Q] =&  e^c \wedge *\left( Q_{ac} \wedge e^b \right) +  e_c \wedge *\left( Q^{bc} \wedge e_a \right), \\
    \overset{(4)}{\Sigma^b{}_a}[Q] =& 2 \delta^b_a *Q  ,\\
    \overset{(5)}{\Sigma^b{}_a}[Q] =& \delta^b_a e^c \wedge *\left( Q_{cd} \wedge e^d \right) + \frac{1}{2} \left[  e_a \wedge *\left( Q \wedge e^b \right) +  e^b \wedge *\left( Q \wedge e_a \right) \right]  .
  \end{align}
   \end{subequations} 
It is worthy to remark that all angular momentum 1-forms are symmetric, $\overset{(i)}{\Sigma_{ab}}[Q] = \overset{(i)}{\Sigma_{ba}}[Q]$. We can eliminate $D\lambda_a$ in the equation (\ref{eq:field-eqn-coframe-stpg}) by taking covariant exterior derivative of (\ref{eq:field-eqn-connec-stpg}),
 \begin{align}
     D\Sigma^b{}_a[Q]   - e^b \wedge D\lambda_a  = 0 
 \end{align}
where we used the results $De^b = T^b =0$ and $D^2\rho^b{}_a = R^b{}_c \wedge \rho^c{}_a - R^c{}_a \wedge \rho^b{}_c =0$ in the symmetric teleparallel geometry. Then, we multiply the result by $\iota_b$ for computing $D\lambda_a$ explicitly
  \begin{align}
       D\lambda_a = \iota_b D\Sigma^b{}_a[Q]  , \label{eq:Dlambdaa-stpg}
  \end{align}
where we used the identities, $\iota_a e^b=\delta^b_a$ and $e^a\wedge \iota_a \alpha = p\alpha$ for any $p$-form $\alpha$. By inserting that into (\ref{eq:field-eqn-coframe-stpg}) we obtain our essential field equation of STPG which is the combination of (\ref{eq:field-eqn-coframe-stpg}) and (\ref{eq:field-eqn-connec-stpg})
  \begin{align}
       G_a[Q,\Lambda] := \iota_b D \Sigma^b{}_a[Q] + \tau_a[Q] + \Lambda *e_a =0 . \label{eq:field-eqn-combined-stpg}
  \end{align}
Accordingly, the new tensor 2-form $ G_a[Q,\Lambda]$ corresponds to the Einstein tensor 2-form $\widetilde{G}_a[\widetilde{R},\Lambda] := -\frac{1}{2} \widetilde{R}^b{}_c \wedge e_{ab}{}^c+ \Lambda *e_a$ for the values (\ref{eq:gr-equiv-values-ci}) with $\kappa=-1$. 

It may be enlightening to compare our STPG model with its counterpart, the so-called metric (Weitzenböck) teleparallel model of gravity (WTPG)\footnote{A gravity model developed in a spacetime with only torsion was called teleparallel model of gravity in the past, but nowadays it is called metric (Weitzenböck) teleparallel model of gravity.}, in which there are much more literature than STPG, see \cite{gonzalez-2012}, \cite{capozziello-2013-jhep}, \cite{saridakis-2016-cqg}, \cite{ren-saridakis-2021-jcap} and the references therein. One generally formulates WTPG in the orthonormal frame. When the orthonormal (or spin) connection 1-form is set to zero, $\omega^a{}_b=0$, as the nonmetricity and the curvature vanish automatically, torsion does not $T^a = de^a \neq 0$ via the definitions (\ref{eq:cartan-ort}). So, zero-spin connection seems enough for Weitzenböck teleparallelism. If one does it from the outset, that is, from the level of Lagrangian, then the developed theory is not invariant under the Lorentz transformations. In this case, the choice of $\omega^a{}_b=0$ means giving up gauge freedom. In other words, it is a gauge fixing. Thus, the variational field equations obtained from that non-invariant Lagrangian will be non-covariant and accordingly the found classes of solutions will be special to that gauge or that frame. As a consequence, diagonal and non-diagonal choices of the coframes give rise to results suffering from frame-dependent artifacts. On the other hand, if one does not substitute zero-spin connection to the Lagrangian of metric teleparallel gravity, then it becomes a Lorentz-invariant and the field equations obtained from it through the independent variations with respect to coframe and connection become covariant. Correspondingly, the solutions to them do not suffer from frame-dependent problems \cite{saridakis-2016-cqg}, \cite{ren-saridakis-2021-jcap}. As seen from our Lagrangians (\ref{eq:lag-stpg}) and (\ref{eq:lag-QQ}), our theory of STPG is Lorentz invariant and the corresponding variational field equations are Lorentz covariant. Therefore, our model is free from artifacts concerning with diagonal or non-diagonal choices of the coframes. For example, the solution class of (\ref{eq:btz-metric}) is valid under the following transformations
  \begin{subequations}
  \begin{align}
      e^0 \to e^{0'} &= \left(1+ \frac{w_0^2}{r^2} \right) e^0 - w_0\frac{f(r)}{r} e^2 \\
      e^1 \to e^{1'} &= e^1 \\
      e^2 \to e^{2'} &= - \frac{w_0}{rf(r)} \left(1+ \frac{w_0^2 + r^2}{r^2} \right) e^0 + \frac{w_0^2 +r^2}{r^2} e^2 
  \end{align}
  \end{subequations}
which are generated by the transformations
 \begin{align}
     t \to t- w_0 \varphi , \qquad r \to r , \qquad \varphi \to \varphi, \qquad w_0 \to -w_0 \, .
 \end{align}

\subsection{The rotating circularly symmetric metric and coincident gauge}

Now, we will solve the field equations (\ref{eq:zero-tors-stpg}), (\ref{eq:zero-curv-stpg}) and (\ref{eq:field-eqn-combined-stpg}). Since BTZ black-hole solution of the Einstein's equation with a negative cosmological constant obtained in the circularly symmetric Riemannian spacetime yields interesting results at both classical and quantum levels and shares several nice properties of the Kerr black hole obtained in 4-dimensional GR, we consider the rotating circularly symmetric metric as well. While searching solution we will prefer to follow the route of the coincident (or natural or inertial) gauge  \cite{Adak2006_1}, \cite{tomi-lavinia2017}. We will write its steps explicitly below. 

\bigskip

\noindent
{\it Step 1.} We make a metric ansatz for circularly symmetric rotating solutions in polar coordinates $(t,r,\varphi)$
 \begin{align}
     ds^2 &= -f^2(r)dt^2 + g^2(r)dr^2 + h^2(r)\left[w(r)dt + d\varphi \right]^2, \label{eq:metric}
 \end{align}
where $f(r),g(r), h(r), w(r)$ are unknown metric functions to be calculated from the field equations.

\bigskip

\noindent
{\it Step 2.} We choose the orthonormal coframe for this metric
 \begin{align} \label{eq:coframe-ansatz}
     e^0 = f(r) dt, \qquad e^1 = g(r) dr , \qquad e^2 = h(r)\left[ w(r) dt + d\varphi \right]
 \end{align}
where $ds^2=-\left(e^0\right)^2 + \left(e^1\right)^2 + \left(e^2\right)^2$. Please, notice that $g$-orthonormal coframe is written in terms of metric functions. 

\bigskip

\noindent
{\it Step 3.} We calculate the dreibeins and its inverse through the relations $e^a=h^a{}_\mu dx^\mu$ and $dx^\mu = h^\mu{}_a e^a$ as
 \begin{align}
    h^a{}_\mu = 
     \begin{bmatrix}
      f & 0 & 0 \\
      0 & g & 0 \\
      w h & 0 & h
     \end{bmatrix}
      \qquad \text{and} \qquad 
     h^\mu{}_a = 
     \begin{bmatrix}
      1/f & 0 & 0 \\
      0 & 1/g & 0 \\
      -w/f & 0 & 1/h 
     \end{bmatrix}
      .
 \end{align}
Again notice that dreibeins are written in terms of metric functions.

\bigskip

\noindent
{\it Step 4.} We write not only the orthonormal coframe, but also the affine connection in terms of the dreibein as $e^a=h^a{}_\mu dx^\mu$ and $\omega^a{}_b = h^a{}_\mu dh^\mu{}_b$. Then, the connection 1-form is calculated
 \begin{align}
     \omega^a{}_b =
     \begin{bmatrix}
        -e^1 f'/fg & 0 & 0 \\
        0 & -e^1g'/g^2 & 0 \\
        -e^1hw'/fg & 0 & -e^1h'/gh
    \end{bmatrix} . \label{eq:full-connect-on1}
 \end{align}
Again notice that affine connection 1-form is written in terms of metric functions. 

\bigskip

\noindent
{\it Step 5.} We compute $T^a=0$, $R^a{}_b=0$ and   
 \begin{align}
     Q_{ab} = \frac{1}{2}(\omega_{ab} + \omega_{ba}) = \begin{bmatrix}
        e^1 f'/fg & 0 & -e^1hw'/2fg \\
        0 & -e^1g'/g^2 & 0 \\
        -e^1hw'/2fg & 0 & -e^1h'/gh
    \end{bmatrix} . \label{eq:Qab-conincident} 
    \end{align}
Again notice that nonmetricity 1-form is written in terms of metric functions.

After calculating $Q_{ab}$ via the coincident gauge in terms of metric functions we assume very well known BTZ metric of the Einstein's equation
 \begin{align} \label{eq:btz-metric}
     h(r)=r, \qquad w(r)=\frac{w_0}{r^2} , \qquad f(r)=\frac{1}{g(r)}= \sqrt{m_0 + \frac{w_0^2}{r^2} - \Lambda r^2} 
 \end{align}
where $m_0$ and $w_0$ are constants. Then, we have checked that the field equation (\ref{eq:field-eqn-combined-stpg}) of STPG is satisfied as long as the following relations for $c_i$s are valid
  \begin{align}
      c_1=-\frac{1}{2}, \qquad c_3=1-c_2, \qquad c_4=\frac{1}{2}, \qquad c_5 = -1 .
  \end{align}
It is worthy to notice that there are four free parameters here, $c_2$, $m_0$, $w_0$, $\Lambda$. Correspondingly, we calculate covariant exterior derivative STPG equation (\ref{eq:field-eqn-combined-stpg}) whether it vanishes or not, because the corresponding operation in general relativity yields zero, $\widetilde{D}\widetilde{G}_a[\widetilde{R},\Lambda]=0$. We saw $D G_a[Q,\Lambda]=0$ like in GR. Furthermore, we calculate covariant derivatives of two equations (\ref{eq:field-eqn-coframe-stpg}) and (\ref{eq:field-eqn-connec-stpg}) separately. Firstly we do it for the coframe equation by noticing $D^2\lambda_a = -R^b{}_a \wedge \lambda_b=0$ and found that $D(\tau_a[Q] + \Lambda *e_a + D\lambda_a)=0$. Secondly we perform it for the connection equation by using (\ref{eq:Dlambdaa-stpg}) and $D^2\rho^b{}_a=0$. As a result we found $ D(\Sigma^b{}_a[Q] + e^b \wedge \lambda_a + D\rho^b{}_a)=0$. Consequently, we have still four free parameter, $c_2$, $m_0$, $w_0$, $\Lambda$, in our hand.

Besides, we found a class of non-GR solution to STPG equations for vanishing cosmological constant, $\Lambda=0$, 
 \begin{align} \label{eq:stpg-non-gr}
      h(r)=r, \qquad w(r)=\frac{w_0}{r^2} , \qquad f(r)=\frac{1}{g(r)}= \sqrt{m_0 + \frac{w_0^2}{r^2} + \frac{m_0^2}{2w_0} r^2} = \frac{m_0 r}{2w_0} + \frac{w_0}{r}
 \end{align}
under the relations 
 \begin{align}
     c_3=-(3c_1 + c_2), \qquad c_4=-c_1, \qquad c_5 = 2c_1 .
 \end{align}
Here there are four free parameters, $c_1$, $c_2$, $m_0$, $w_0$. Since GR-equivalent values (\ref{eq:gr-equiv-values-ci}) do not satisfy the constraint $c_3=-(3c_1+c_2)$ we call it as non-GR solution. While searching exact solutions we heavily use the computer algebra system REDUCE \cite{hearn-2004} and its exterior algebra package EXCALC \cite{schrufer-2004}. 

Finally, we remark on the singularities and the horizons of the above solutions. In this context, the first step is to find at which $r$ do the metric functions become zero or infinity. However, since these singularities could correspond to coordinate singularities, the usual process is to compute various invariants. Then, if these invariants diverge at one point, they will do it independently of the specific coordinate system. It is worthy to notice that the opposite is not true, i.e., the finiteness of an invariant is not a proof that there is not a physical singularity there, e.g, BTZ black hole of GR. Since in STPG all invariants generated from nonmetricity are cast in the Lagrangian (\ref{eq:lag-QQ}), we will compute 0-form $*L[Q^2]$. Meanwhile, we will calculate the following 0-form $*K[\widetilde{R}]$ written in terms the scalar invariants of the Riemannian curvature 2-form,   
 \begin{align}
    K[\widetilde{R}] =  l_1 \widetilde{R}^{ab} \wedge *e_{ab} +  l_2 (\iota_a\widetilde{R}^{ab}) \wedge * (\iota^c\widetilde{R}_{cb}) + l_3 \widetilde{R}^{ab} \wedge *\widetilde{R}_{ab} 
 \end{align} 
where $l_1, l_2, l_3$ are arbitrary constants to be used for tracing the invariants, $\widetilde{R}^{ab} \wedge *e_{ab}$ is the Riemannian Einstein-Hilbert 3-form containing the Riemannian curvature scalar, $\iota_a\widetilde{R}^{ab}$ is the Riemannian Ricci curvature 1-form and the last term yields the Kretschmann scalar. Thus, we are able to compare two results and deduce some insights. For the solution (\ref{eq:btz-metric}), while three singular points are obtained from the metric functions 
 \begin{align}
     r_0=0, \qquad r_\pm= \sqrt{ \frac{m_0}{2\Lambda} \left(  1 \pm \sqrt{1- \frac{4\Lambda w_0^2}{m_0^2}} \right)}
 \end{align}
the both invariants are finite everywhere  
      \begin{align}
          *L[Q^2] = - \Lambda , \qquad      *K[\widetilde{R}] = -6 l_1 \Lambda - 12 l_2 \Lambda^2 - 13 l_3 \Lambda^2 .
      \end{align}
As seen, $w_0^2 \leq m_0^2/4\Lambda$ must be valid for the existence of $r_\pm$ singularities. Thus, we can deduce that our STPG solution (\ref{eq:btz-metric}) has the same singularity and horizon structures as GR.  We repeat the similar calculations for the solution (\ref{eq:stpg-non-gr}) and arrive at 
 \begin{align}
     r_0=0 , \qquad r= \sqrt{\frac{-2w_0^2}{m_0}} , \qquad 
     *L[Q^2] = 0 , \qquad *K[\widetilde{R}] = \frac{12l_1 m_0^2 w_0^2 - 6l_2 m_0^4 - 3l_3 m_0^4}{8w_0^4}.
 \end{align}
This time, for existence of an outer singularity (horizon) it must be $m_0 < 0$. Then, we conclude that this non-GR solution may have black hole properties as well. In order to understand singularity, horizon and black hole structures of symmetric teleparallel gravity models in a deeper level more rigorous analysis similar to the Raychaudhuri and optical equations for null geodesic congruences in GR is needed \cite{yang-harko-2021}.

\subsection{Minimal coupling of Maxwell field to STPG theory}

As the second stage we include electromagnetic field in the game. We couple minimally the electromagnetic field to STPG by adding the Maxwell Lagrangian to our STPG Lagrangian (\ref{eq:lag-stpg}) 
 \begin{align}
     L_2[Q,A] = L_1[Q] + dA \wedge *dA \label{eq:lagr-stpg-maxw}
 \end{align}
where $A$ is the electromagnetic potential 1-form. This Lagrangian is invariant under $SO(1,2) \otimes U(1)$ group. The variational field equations are obtained as
   \begin{subequations} \label{eq:equations-stpg-maxw}
 \begin{align}
     T^a &= 0 , \label{eq:zero-tors-stpg-maxw}\\
     R^a{}_b &= 0, \label{eq:zero-curv-stpg-maxw} \\
      \tau_a[Q] + \Lambda *e_a + \tau_a[A] + D\lambda_a &= 0 , \label{eq:field-eqn-coframe-stpg-maxw} \\
      \Sigma^b{}_a[Q] +  e^b \wedge \lambda_a  +D\rho^b{}_a &= 0, \label{eq:field-eqn-connec-stpg-maxw} \\
      d *dA &=0 \label{eq:maxwell-stpg},
 \end{align}
  \end{subequations}
where $\tau_a[Q]$ is given by (\ref{eq:tauQ-stpg}), $\Sigma^b{}_a[Q]$ by (\ref{eq:SigmaQQ-stpg}) and the energy-momentum 2-form of electromagnetic field is
 \begin{align}
     \tau_a[A]= -\big( \iota_a dA\big) \wedge *dA + dA \wedge \big( \iota_a *dA \big) . \label{eq:ener-momen-maxwell}
 \end{align}
The equation (\ref{eq:maxwell-stpg}) is the Maxwell equation. Again, we combine the coframe and connection equations by repeating the previous calculation and obtain our STPG equation
 \begin{align}
       \iota_b D \Sigma^b{}_a[Q] + \tau_a[Q] + \tau_a[A] + \Lambda *e_a =0 . \label{eq:field-eqn-combined-stpg-maxw}
  \end{align}
Together with the metric ansatz (\ref{eq:coframe-ansatz}) we assume the electromagnetic potential 1-form 
 \begin{align}
          A = E(r) dt + B(r) d\varphi, \label{eq:maxwell-assumption1}
 \end{align}
where $E(r)$ and $B(r)$ are new unknown functions aside from the metric functions to be calculated from the field equations. With help of the literature on spinning charged solutions to three-dimensional Einstein-Maxwell theory, e.g.\cite{clement-1996},\cite{martinez2000},\cite{dereli-obukhov-2000}, we have considered the following configuration
  \begin{subequations} \label{eq:solution-stpg-maxwell}
      \begin{align}
      f(r) &= \frac{r}{h(r)} \sqrt{\frac{r^2}{l^2} - \frac{l_0^2q_0^2}{l^2} \ln\left(\frac{r}{r_0}\right)} ,\\
     g(r) &= \frac{r}{h(r) f(r)},\\
     h(r) &= \sqrt{r^2 + w_0^2 q_0^2 \ln\left(\frac{r}{r_0}\right)} , \\
    w(r) &= -  \frac{w_0 q_0^2}{[h(r)]^2} \ln\left(\frac{r}{r_0}\right), \\
      E(r) = - \frac{B(r)}{w_0} &= - k_0 q_0 \ln\left(\frac{r}{r_0}\right),
       \end{align}
   \end{subequations}
where $l$, $l_0$, $r_0$, $q_0$, $w_0$, $k_0$ are free parameters. Under the following relations among the free parameters and coupling constants
 \begin{align}
     c_1= -k_0^2, \qquad c_3= 2k_0^2-c_2, \qquad c_4=k_0^2, \qquad c_5=-2k_0^2, \nonumber\\
     \Lambda=-\frac{2k_0^2}{l^2}, \quad l_0^2 = l^2-w_0^2,
 \end{align}
the functions (\ref{eq:solution-stpg-maxwell}) are a class of solution to our field equations (\ref{eq:equations-stpg-maxw}). Please notice that $c_2$, $w_0$, $q_0$, $r_0$, $l$, $k_0$ are six free parameters. We also calculated the covariant derivatives of the coframe, connection and combined equations respectively,
   \begin{subequations} 
 \begin{align}
          D\left( \tau_a[Q] + \Lambda *e_a + \tau_a[A] + D\lambda_a \right) &= 0 , \\
      D\left(\Sigma^b{}_a[Q] +  e^b \wedge \lambda_a  +D\rho^b{}_a\right) &= 0, \\
       D\left(\iota_b D \Sigma^b{}_a[Q] + \tau_a[Q] + \tau_a[A] + \Lambda *e_a\right) &=0 . 
 \end{align}
  \end{subequations}
Furthermore, we want to remark that although the covariant derivative of the electromagnetic energy-momentum with respect to the full connection do not vanish, $D \tau_a[A] \neq 0$, it is zero with respect to the Levi-Civita (Riemannian) connection, $\widetilde{D} \tau_a[A] = 0$.  That can be explained through the decomposition (\ref{eq:connec-decom}) which gives $ D\tau_a[A] = \widetilde{D}\tau_a[A] - \mathrm{L}^b{}_a \wedge \tau_b[A]$. If someone persistently asks for $D\tau_a[A]=0$, it becomes possible by setting $l^2=w_0^2$ (or $l_0=0$). In this case, the unknown functions turn out to be
  \begin{align} \label{eq:solution2-stpg-maxwell-scalar}
     f(r)=\frac{r^2}{l h(r)}, \quad g(r)= \frac{l}{r}, \quad h(r) = \sqrt{r^2 + l^2 q_0^2 \ln\left(\frac{r}{r_0}\right)} ,\nonumber \\
     w(r) = - \frac{l q_0^2}{[h(r)]^2} \ln\left(\frac{r}{r_0}\right), \quad  E(r) = - \frac{B(r)}{l} = - k_0 q_0 \ln\left(\frac{r}{r_0}\right).
      \end{align}
Then, the number of free parameters reduces to five.

Now, in order to proceed to the singularities and horizons investigation along the lines at the end of the previous subsection, we have to solve the equation $r^2 - l_0^2q_0^2 \ln\left(\frac{r}{r_0}\right) =0$ coming from the solution (\ref{eq:solution-stpg-maxwell}) in order to obtain singular points apart from $r=0$. However, we can clearly see that in general this is a transcendental equation, whose roots cannot be obtained analytically. Thus, by plotting the graphs of two functions $y_1(r)=r^2$ and $y_2(r)=l_0^2q_0^2 \ln\left(\frac{r}{r_0}\right)$ on the same system of axes and then searching their intersection points one can see that three cases are possible in general. There may be two intersections (singular points) at most or only one intersection (singular point) or none. They depend on values of constants, $l_0$ and $q_0$. We also calculated the related invariants, $*L[Q^2]$ and $*K[\widetilde{R}]$, and saw that they have singularity only at $r=0$. Thus, we conclude that when there is at least one outer singular point shielding the one at origin, that structure may define a black hole. A discussion on the singularity structure of the solution (\ref{eq:solution2-stpg-maxwell-scalar}) will be written at end of the next subsection.

\subsection{Nonminimal coupling of scalar field to the Maxwell-STPG theory}

As the third and final stage we couple nonminimally a scalar field, $\phi$, to our Maxwell-STPG theory (\ref{eq:lagr-stpg-maxw}) as follows
    \begin{align} \label{eq:lagr-stpg-maxw-scalar}
     L_3[Q,A,\phi] =& \phi L[Q^2] + \phi^{-1} d\phi \wedge * d\phi + c_6 Q \wedge *d\phi + c_7 (Q_{ab} \wedge e^b) \wedge *(d\phi \wedge e^a)   \nonumber \\
     & \qquad + \Lambda \phi^3 *1 + \phi^{-1} dA \wedge *dA + T^a \wedge \lambda_a + R^a{}_b \wedge \rho^b{}_a ,
 \end{align}
where $c_6$ and $c_7$ are new coupling constants. For $\phi=1$ the Lagrangian (\ref{eq:lagr-stpg-maxw-scalar}) reduces to Maxwell-STPG Lagrangian (\ref{eq:lagr-stpg-maxw}). $SO(1,2)\otimes U(1)$-invariance of Lagrangian (\ref{eq:lagr-stpg-maxw-scalar}) is explicit, but is also invariant under the Weyl transformation (\ref{eq:weyl}) together with the relations among $c_i$s,
   \begin{subequations} \label{eq:c-s-weyl-invar}
     \begin{align}
      2c_2 + 2c_3 + 3c_5 + c_7 &=0 , \\
      2c_1 + 2c_2 + 2c_3 + 6c_4 + 5c_5 + c_6 + c_7 &=0 , \\
      1 + 3c_1 + 2c_2 + 2c_3 + 9c_4 + 6c_5 + 3c_6 + 2c_7 &=0, \\
      2 + 3c_6 + 2c_7 &=0 .
     \end{align}
   \end{subequations}
Although four equations appear, three are linearly independent. It may be worthwhile to point that by defining the Weyl-covariant exterior derivative of scalar field 
  \begin{align}
     \mathcal{D}\phi = d\phi + \left[ k_6 \,  \eta^{ab}Q_{ab} + k_7 \, (\iota_aQ^{ab})e_b \right]\phi 
 \end{align}
we can rewrite the Lagrangian (\ref{eq:lagr-stpg-maxw-scalar}) in a tidier form as
     \begin{align} \label{eq:lag2}
     {L'}_3[Q,A,\phi] =& \phi L'[Q^2] + \phi^{-1} \mathcal{D}\phi \wedge * \mathcal{D}\phi + \Lambda \phi^3 *1    + \phi^{-1} dA \wedge *dA \nonumber \\
     & \qquad + T^a \wedge \lambda_a + R^a{}_b \wedge \rho^b{}_a ,
 \end{align}
where 
  \begin{align} 
     L'[Q^2] =& k_1  Q_{ab} \wedge *Q^{ab} + k_2 \left(Q_{ab} \wedge e^b\right) \wedge *\left(Q^{ac} \wedge e_c\right) + k_3 \left(Q_{ab} \wedge e_c \right) \wedge *\left(Q^{ac} \wedge e^b\right)  \nonumber \\
     & \qquad + k_4 Q \wedge *Q   + k_5 (Q \wedge e^b) \wedge * (Q_{ab} \wedge e^a) .
 \end{align}
Here $k_i$, $i=1,2, \cdots , 7$, are new coupling constants. We showed that Lagrangian (\ref{eq:lagr-stpg-maxw-scalar}) and Lagrangian (\ref{eq:lag2}) are equivalent under the following relations among the coupling constants,
     \begin{align}
     &c_1 = k_1 + k_7^2 , \qquad  c_2 = k_2 ,\qquad   c_3 = k_3 -k_7^2 , \qquad    c_4 = k_4 + k_6^2 + 2k_6k_7 ,\\
     & \qquad \qquad c_5 = k_5 -2k_6k_7 , \qquad   c_6 = 2(k_6 + k_7) , \qquad  c_7 = -2k_7  . \nonumber  
    \end{align}
We have also checked that the algebraic equations (\ref{eq:c-s-weyl-invar}) for Weyl invariance are consistent with (56a)-(56c) of Ref.\cite{koivisto-Iosifidis2018} with $n=3$. 

Thus, the variations $\delta \lambda_a$, $\delta \rho^b{}_a$, $\delta \phi$, $\delta A$, $\delta e^a$ and $\delta \omega^a{}_b$ yield the following field equations, respectively,
 \begin{subequations} \label{eq:solution-stpg-maxwell-scalar}
 \begin{align}
     T^a &= 0 , \label{eq:zero-tors}\\
     R^a{}_b &= 0, \label{eq:zero-curv} \\
    L[Q^2]   - 2 \phi^{-1} d  *d\phi + \phi^{-2} d\phi \wedge * d\phi + 3\Lambda \phi^2 *1 - \phi^{-2} dA \wedge *dA \qquad \qquad   & \nonumber \\
       - c_6 \, d*Q - c_7 \, d[e^a \wedge *(Q_{ab} \wedge e^b)]  &=0 , \label{eq:field-eqn-scalar}\\ 
      d\big( \phi^{-1} *dA \big) &=0, \label{eq:field-eqn-maxwell}\\
      \tau_a[Q] + \tau_a[\phi] + \tau_a[A] + D\lambda_a &= 0 , \label{eq:field-eqn-coframe1} \\
      \Sigma^b{}_a[Q] + \Sigma^b{}_a[\phi]  + e^b \wedge \lambda_a + D\rho^b{}_a &= 0. \label{eq:field-eqn-connec1}
 \end{align}
  \end{subequations}
Energy-momentum 2-form of nonmetricity is 
  \begin{align} 
 \tau_a[Q]= \phi \sum_{i=1}^5 c_i \overset{(i)}{\tau_a}[Q] 
 \end{align}
where $\overset{(i)}{\tau_a}[Q]$ are given by equations (\ref{eq:enr-mom-nonmet-sqred}). Angular (hyper) momentum 2-form of nonmetricity is 
  \begin{align}  
 \Sigma^b{}_a[Q]= \phi \sum_{i=1}^5 c_i \overset{(i)}{\Sigma^b{}_a}[Q]
 \end{align}
where $\overset{(i)}{\Sigma^b{}_a}[Q]$ are given by the equations  (\ref{eq:ang-mom-nonmet-sqred}). Energy-momentum 2-form of electromagnetic field is by the equation (\ref{eq:ener-momen-maxwell}). 
  \begin{align}
      \tau_a[A]= - \phi^{-1} \left[ \big( \iota_a dA\big) \wedge *dA - dA \wedge \big( \iota_a *dA \big) \right] . \label{eq:ener-momen-maxwell-scal}
  \end{align}
Energy-momentum 2-form of the scalar field is
 \begin{align}
    \tau_a[\phi] =& - \phi^{-1} [ \big(\iota_a d\phi \big) \wedge *d\phi + d\phi \wedge \big(\iota_a *d\phi \big)  ] -c_6 [ (\iota_ad\phi) \wedge *Q + Q \wedge (\iota_a *d\phi) ]  \nonumber \\
       &- c_7[  \iota_a(d\phi \wedge e^b) \wedge *(Q_{bc} \wedge e^c) - (Q_{bc} \wedge e^c) \wedge \iota_a*(d\phi \wedge e^b) ]  \nonumber \\
       &- c_7 [ Q_{ab} \wedge *(d\phi \wedge e^b) + d\phi \wedge *(Q_{ab} \wedge e^b)] + \Lambda \phi^3 *e_a .
 \end{align}
Angular momentum 2-form of scalar field is
  \begin{align}
     \Sigma^b{}_a[\phi] =& c_6 \delta^b_a  *d\phi +  \frac{c_7}{2}\bigg[ e^b \wedge *(d\phi \wedge e_a) + e_a \wedge *(d\phi \wedge e^b) \bigg] . 
  \end{align}
Because of the nonminimal couplings angular momentum of scalar field appears. It is worthy to remark that all angular momentum 1-forms are symmetric, $\Sigma_{ab} = \Sigma_{ba}$. Again as done earlier we combine the coframe and connection equations  as follows
  \begin{align}
       \iota_b D\big( \Sigma^b{}_a[Q] + \Sigma^b{}_a[\phi]\big) +  \tau_a[Q]  + \tau_a[A] + \tau_a[\phi] = 0 . \label{eq:field-eqn-combined}
  \end{align}
Although the field equations we must solve are (\ref{eq:zero-tors}), (\ref{eq:zero-curv}), (\ref{eq:field-eqn-scalar}), (\ref{eq:field-eqn-maxwell}) and (\ref{eq:field-eqn-combined}), since the zero-torsion and zero-curvature equations are satisfied automatically by the coincident gauge, we effectively have to solve the equations (\ref{eq:field-eqn-scalar}), (\ref{eq:field-eqn-maxwell}) and (\ref{eq:field-eqn-combined}).

Before attempting to find a solution, we want to remark on the Maxwell equation (\ref{eq:field-eqn-maxwell}). The effects of non-minimal coupling of electromagnetic fields to other fields can be encoded into the definition a constitutive tensor. For electromagnetic 2-form, $F=dA$, in an arbitrary medium Maxwell's equations could be expressed  
 \begin{align}
     dF =0 \qquad \text{and} \qquad d*G = J
 \end{align}
where $G$ is called the excitation 2-form and $J$ is the source electric current 2-form (3-form in 4-dimensions). To close the system we need electromagnetic constitutive relations among $G,J,F$. In our case $J=0$ and we assume a simple linear constitutive relation $G = \mathcal{Z}(F)$ where $\mathcal{Z}$ is $(2,2)$-type constitutive tensor. For our case, it is
 \begin{align}
     G = \phi^{-1} F .
 \end{align}
 
As the final job in this section we want to find a class of circularly symmetric rotating solution to the equations (\ref{eq:field-eqn-scalar}), (\ref{eq:field-eqn-maxwell}) and (\ref{eq:field-eqn-combined}) by assuming the equations (\ref{eq:Qab-conincident}), (\ref{eq:maxwell-assumption1}) and $\phi=\phi(r)$. Firstly, we found that the solution (\ref{eq:solution-stpg-maxwell}) is also a solution to the equations (\ref{eq:solution-stpg-maxwell-scalar}) for $\phi(r)=1$ under the conditions
   \begin{align}
     c_1= -k_0^2, \qquad c_3= 2k_0^2-c_2, \qquad c_4=k_0^2, \qquad c_5=-2k_0^2, \nonumber\\
     c_6= 0, \quad c_7= 2k_0^2, \quad \Lambda=-\frac{2k_0^2}{l^2}, \quad l_0^2 = l^2-w_0^2.
 \end{align}
Here there are six free parameters, $c_2, k_0, l, r_0,w_0,q_0$. These $c_i$ values do not satisfy the scale invariance conditions (\ref{eq:c-s-weyl-invar}). Besides, we checked $D\text{(\ref{eq:field-eqn-combined})}=0$, $D\text{(\ref{eq:field-eqn-coframe1})}=0$,  $D\text{(\ref{eq:field-eqn-connec1})}=0$, $D\tau_a[A] \neq 0$ and $\widetilde{D}\tau_a[A] = 0$. When $ w_0 = l$ (or $l_0=0$), in which case the number of free parameters reduces to five, it becomes $D\tau_a[A] =0$ and (\ref{eq:solution2-stpg-maxwell-scalar}) is the new solution set. In this case the scale invariance conditions (\ref{eq:c-s-weyl-invar}) are still not satisfied.

Secondly, by virtue of the Weyl transformation (\ref{eq:weyl}) we write down the orthonormal coframe and affine connection together with $\phi(r)= e^{-\psi(r)}$ as
 \begin{align} 
     e^0 &= e^{\psi(r)} f(r) dt, \qquad e^1 = e^{\psi(r)} g(r) dr , \qquad e^2 = e^{\psi(r)} h(r)\left[ w(r) dt + d\varphi \right], \label{eq:coframe-ansatz-weyl} \\
     \omega^a{}_b &= 
     \begin{bmatrix}
        -e^1 (f'+f\psi')/(fge^\psi) & 0 & 0 \\
        0 & -e^1(g'+g\psi')/(g^2e^\psi) & 0 \\
        -e^1hw'/(fge^\psi) & 0 & -e^1(h'+h\psi')/(ghe^\psi)
    \end{bmatrix}  \label{eq:full-connect-weyl}
 \end{align}
where $\psi(r)$ is called the scale (Weyl) function. This connection (\ref{eq:full-connect-weyl}) is the affine connection derived from the orthonormal coframe (\ref{eq:coframe-ansatz-weyl}) through the coincident gauge recipe. In this case the functions (\ref{eq:solution2-stpg-maxwell-scalar}) are solutions under the constraints,
  \begin{align}
     c_3= k_0^2-c_1-c_2, \qquad c_4=\frac{1- 7c_1-2k_0^2}{3}, \qquad c_5=\frac{10c_1+2k_0^2-1}{3}, \nonumber\\
     c_6= \frac{16c_1+8k_0^2-4}{3}, \quad c_7=1-8c_1-4k_0^2, \quad \Lambda=\frac{16c_1 +8k_0^2-1}{3l^2}, \quad w_0=l.
 \end{align}
Now, $c_1, c_2, k_0, l, r_0, q_0, \psi(r)$ are arbitrary. This configuration satisfies the scale invariance conditions (\ref{eq:c-s-weyl-invar}). Besides, we verified $D\text{(\ref{eq:field-eqn-combined})}=0$, $D\text{(\ref{eq:field-eqn-coframe1})}=0$,  $D\text{(\ref{eq:field-eqn-connec1})}=0$, $D\tau_a[A] = 0$ and $\widetilde{D}\tau_a[A] = 0$. 

For singularity analysis of this last solution we look at the equations (\ref{eq:solution2-stpg-maxwell-scalar}). Those functions have only one singularity at $r=0$, no horizon meaning that they are naked singularities. But, here our metric functions are obtained by multiplying them by $e^{\psi(r)}$ and the function $\psi(r)$ is arbitrary. We again calculated $*L[Q^2]$ and $*K[\widetilde{R}]$ and saw that they could posses singular points depending on $\psi(r)$.  Consequently, depended on the behaviour of $\psi(r)$ there may be outer singularities apart from the one at origin.

\section{Discussion}

In $1+2$ dimensions Einstein's general theory of relativity does not have dynamical (propagating) degrees of freedom in vacuum, therefore there is a wide literature on modified theories of three dimensional gravity. In this work we too investigate an alternative model of gravity by going to a non-Riemannian spacetime, more precisely the symmetric teleparallel spacetime defined by just the nonmetricity tensor. We formulated our theory by a Lagrangian 3-form and then obtained the field equations by independent variations. By adhering the coincident gauge which was applied explicitly step by step we warranted the conditions of zero-torsion and zero-curvature and could obtain the orthonormal coframe and the nonmetricity in terms of only a metric ansatz. We visited three models, adjacently, i. STPG, ii. minimal Maxwell-STPG, iii. non-minimal scalar-Maxwell-STPG. We founded some classes of exact solutions for each and give some discussions on the singularity and horizon structures. As far as we know all these results are new for the symmetric teleparallel gravity literature. Furthermore, although GR is not a dynamical theory in 3 dimensions, we showed that three dimensional STPG is dynamic. In fact, we know that STPG has dynamical degrees of freedom even in 2 dimensions \cite{adak2008}, but GR is trivial. Thus, one can deduce that STPG has much richer structure than GR.

As with future perspective, we think that mathematical results and techniques developed in this work may find opportunity of applicability in material science concerning the crystal defects because there is a wide literature on the non-Riemannian interpretations relating torsion, curvature and nonmetricity with densities of dislocations, disclinations and metric anomalies, respectively,  \cite{dereli-vercin-1987}-\cite{roychowdhury-gupta-2017} and the references therein. Especially, photonic crystal field seems a suitable avenue for 1+2 dimensional scale invariant symmetric teleparallel geometry. Let us explain why. In photonic crystal researches, in which the aim is to manipulate behaviour of electromagnetic wave (photon) by forming certain defects in crystal lattice, firstly computational studies are performed numerically in a scale independent way. After meaningful results, the associated experiments are performed in micrometer scale laboratories. Since Maxwell's electromagnetic theory is scale invariant, measurements done micrometer scale will be valid in nanometer scale as well. Although 3 space dimensional photonic crystal structures are possible, in practice 2 space dimensional ones are studied \cite{yuksel-berberoglu-2022a}-\cite{oguz-karakilinc-2022b}. Consequently, while there are crystal defects in 2 space dimensions, Maxwell field and scale invariance in theoretical photonic crystal researches, there are nonmetricity, Maxwell field and a scalar field for scale invariance in our model in this paper. Therefore, we conclude that the mathematical techniques and results obtained in this work may lead to new insights in the photonic crystal studies. It is among our future projects.





\section*{Acknowledgements}

One of the authors (MA) stays at Istanbul Technical University (ITU) via a sabbatical leave and thanks the Department of Physics Engineering, ITU for warm hospitality. C.P. and M.A. were supported via the project number 2022FEBE032 by the Scientific Research Coordination Unit of Pamukkale University. One of the authors (C.P.) thanks TUBITAK (Scientific and Technical Research Council of Turkey) for a grant through TUBITAK 2214-A that makes his stay in the Estonia possible and the Institute of Physics, University of Tartu for warm hospitality. We are grateful to Tomi S. Koivisto for fruitful discussions and the anonymous referee for useful comments.

\end{document}